# Identifying and Predicting Parkinson’s Disease Subtypes through Trajectory Clustering via Bipartite Networks

Sanjukta Krishnagopal[1]*, Rainer Von Coelln[2], Lisa M. Shulman[2], Michelle Girvan[1]

1 University of Maryland College Park, College Park, MD 20472, USA.

2 University of Maryland School of Medicine, Baltimore, MD 21201, USA.

## Abstract

Parkinson’s disease (PD) is a common neurodegenerative disease with a high degree of heterogeneity in its clinical features, rate of progression, and change of variables over time. In this work, we present a novel data-driven, network-based Trajectory Profile Clustering (TPC) algorithm for 1) identification of PD subtypes and 2) early prediction of disease progression in individual patients. Our subtype identification is based not only on PD variables, but also on their complex patterns of progression, providing a useful tool for the analysis of large heterogenous, longitudinal data. Specifically, we cluster patients based on the similarity of their trajectories through a time series of bipartite networks connecting patients to demographic, clinical, and genetic variables. We apply this approach to demographic and clinical data from the Parkinson’s Progression Markers Initiative (PPMI) dataset and identify 3 patient clusters, consistent with 3 distinct PD subtypes, each with a characteristic variable progression profile. Additionally, TPC predicts an individual patient’s subtype and future disease trajectory, based on baseline assessments. Application of our approach resulted in 74% accurate subtype prediction in year 4 in a test/validation cohort. Furthermore, we show that genetic variability can be integrated seamlessly in our TPC approach. In summary, using PD as a model for chronic progressive diseases, we show that TPC leverages high-dimensional longitudinal datasets for subtype identification and early prediction of individual disease subtype. We anticipate this approach will be broadly applicable to multidimensional longitudinal datasets in diverse chronic diseases.

## Introduction

Parkinson’s disease (PD) is the most common neurodegenerative movement disorder, affecting an estimated 7-10 million people worldwide[1]. The cause of PD is unknown, and the disease course is variable with age of onset and rate of progression differing across the population[2]. Furthermore, the clinical presentation is variable, with a broad range of possible motor and non-motor symptoms[3]. Based on these differences, multiple PD subtypes have been proposed, based on clinical intuition or unbiased data-driven approaches like cluster analysis[4]. There is currently no consensus on which subtypes are biologically valid and clinically relevant, and the best approach for identifying such subtypes remains elusive[5]. Lack of integration of the longitudinal data for a large number of variables is a limitation of existing approaches.

Network medicine[6-10] offers a promising approach for untangling the complexities due to multiple influences on disease via analysis of interconnections within data. For example, studies of the human disease network (i.e. the 'diseaseome')[10], in which diseases are linked if they share one or more associated genes, are useful for identifying disease pathways and predicting other disease-related genetic variants[8]. With a few exceptions, most network medicine studies have focused on biomolecular data[10-13] rather than the complexities of clinical phenotypic assessments, and disease subtyping based on disease progression patterns is relatively unexplored[14,15]. Additionally, a large number of genetic variants have been identified as risk factors to develop PD. Recently, evidence has emerged that the same genetic risk variants also determine certain clinical features of the disease[16,17]. Genetic factors likely play a major role in determining PD subtypes, however, little work has been done to incorporate genetic data into data-driven algorithms to identify subtypes. One related recent study described inclusion of genetic data to predict the annual rate of change in combined scores from the Movement Disorder Society-Unified Parkinson's Disease Rating Scale[16].

Technological advances in data processing and storage capacity have enabled development of large clinical datasets, containing longitudinal clinical and biological data. In this work we use data from the Michael J. Fox Foundation's Parkinson's Progression Markers Initiative (PPMI), a worldwide study to establish a comprehensive set of clinical, imaging and genetic data (http://www.ppmi-info.org). Such datasets require sophisticated data-driven approaches for effective extraction and analysis of clinically relevant information. Data-driven methods are typically applied to diseases in two ways: *disease-specific*, i.e., identifying disease subtypes and variable progression patterns from large scale patient data, and *patient-specific*, i.e., predicting disease subtype and trajectory in the individual patient based on their data. Our work incorporates both these perspectives and presents a network science method that not only identifies disease subtypes using diverse types of patient data (e.g., genetic and clinical variables), but also is predictive. We present our results based on a (PD dataset, however this method is easily applied to other chronic medical conditions.

To provide an intuitive data-driven solution that is both disease- and patient-centric, we develop the Trajectory Profile Clustering (TPC) algorithm to identify PD subtypes through similarities in patterns of progression. With this approach, we identify three PD subtypes. Additionally, we demonstrate the predictive ability of our algorithm on a test/validation cohort of new patients. Our method correctly predicts the individual PD subtype four years in advance, in 74% of the population. We also explore inclusion of four PD genetic variants in our approach, to demonstrate its capacity to simultaneously incorporate clinical, demographic, and genetic information.

This work is aimed at bridging the gap between the computational methodologies developed by network and data scientists and the clinical experience of health professionals.

## Results

### Description of Data

| Domain | Scale/Variable |
|---|---|
| **Demographics** | |
| Gender | Gender |
| Age | Age |
| **General PD Severity** | |
| MDS-UPDRS1 | Movement Disorders Society-Unified Parkinson's Disease Rating Scale (MDS-UPDRS) - Part 1 |
| MDS-UPDRS2 | Movement Disorders Society-Unified Parkinson's Disease Rating Scale (MDS-UPDRS) - Part 2 |
| MDS-UPDRS3 | Movement Disorders Society-Unified Parkinson's Disease Rating Scale (MDS-UPDRS) - Part 3 |
| T-MDS-UPDRS | Total Movement Disorders Society-Unified Parkinson's Disease Rating Scale (MDS-UPDRS) |
| **Cognitive** | |
| JOLO | Benton Judgement of Line Orientation |
| SDM | Symbol Digit Modalities Test |
| MoCA | Montreal Cognitive Assessment |
| HVLT | Hopkins Verbal Learning Test |
| LNS | Letter Number Sequencing |
| SFT | Semantic Fluency Test |
| **Disability** | |
| SEADL | Schwab and England Activities of Daily Living |
| **Sleep** | |
| RBDQ | Rapid Eye Movement Sleep Behavior Disorder Questionnaire |
| ESS | Epworth Sleepiness Scale |
| **Autonomic** | |
| SCOPA-AUT | Scales for Outcomes in Parkinson's Disease - Autonomic |
| **Mental Health** | |
| GDS | Geriatric Depression Scale |
| STAI | State -Trait Anxiety Inventory |
| **Genetic Risk Loci** | |
| G1 | rs11060180 |
| G2 | rs6430538 |
| G3 | rs823118 |
| G4 | rs356181 |

Table 1: PPMI Data used in this study include two demographic variables, outcome variables from six clinical domains, and four genetic single nucleotide polymorphisms. Additional description of the measures found at http://www.ppmi-info.org.

Data used in the preparation of this article were obtained from the Parkinson's Progression Markers Initiative (PPMI) database (www.ppmi-info.org/data). Of the 430 patients at baseline in this dataset, 314 patients remained in year 4. Once patients with incomplete data were excluded, 198 patients remained for our analysis. Twenty percent of the sample population (n=38) was kept as a test/validation dataset. The remainder of the patients (n=160) form the training dataset that was used in the algorithm to identify PD subtypes. The data included demographics, clinical outcome variables from six clinical domains (General PD Severity, Disability, Cognition, Autonomic Function, Sleep, and Mental Health) and 4 PD genetic variants (Table 1).

### Trajectory Profile Clustering Algorithm

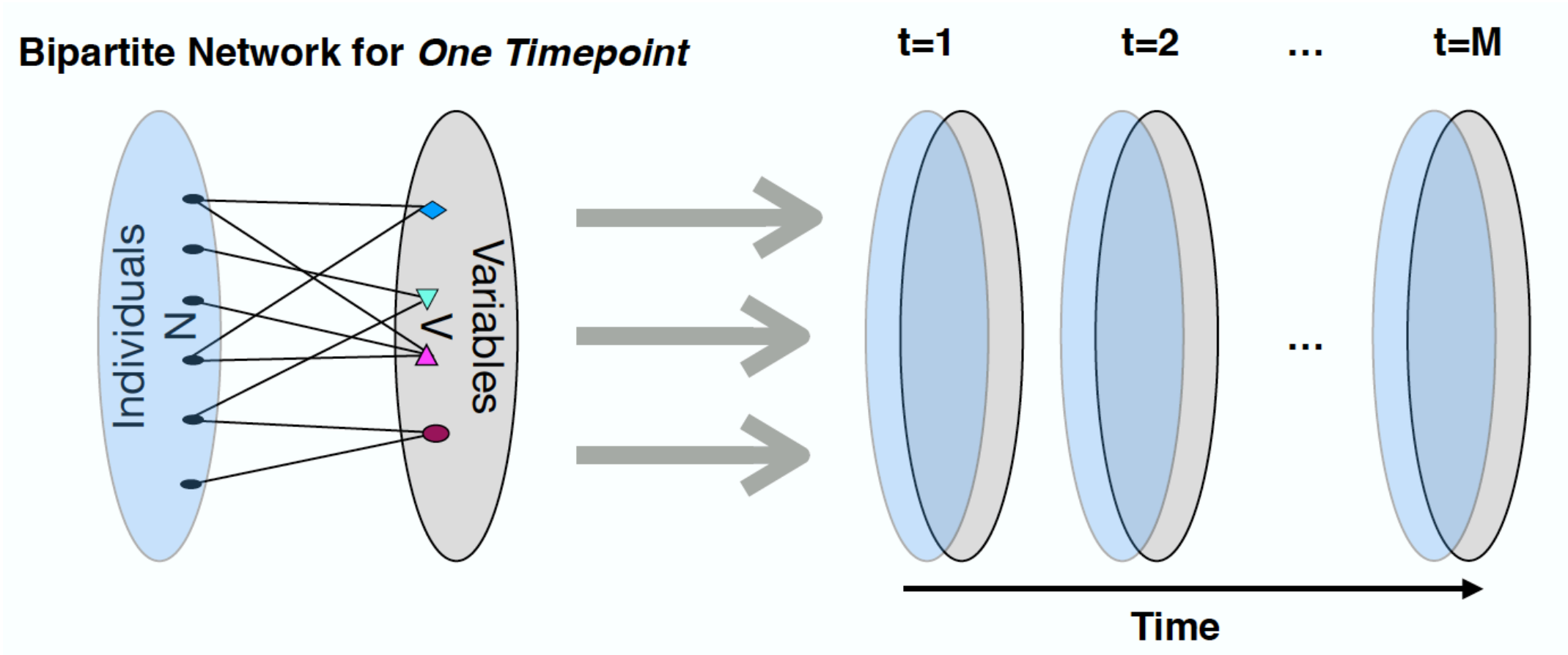

Fig. 1: (left) An illustration of an individual-variable bipartite graph at one timestep. (right) Set of bipartite graphs across time.

Our Trajectory Profile Clustering (TPC) Algorithm is designed to group together patients based on the similarities of their disease trajectories. The algorithm proceeds as follows:

1. *Create bipartite networks connecting individuals to variables:* At a single point in time $t$ (e.g., baseline, year 1, year 2, etc.) we construct an $N \times V$ bipartite graph indicating the connections between individuals and disease variables, where $N$ is the number of individuals in the population and $V$ is the total number of variables measured, as illustrated in Fig. 1. For $M$ time points, we can represent the set of these bipartite graphs as an $N \times V \times M$ multidimensional array, where $X_{ivt}$ gives the value of individual $i$'s disease variable $v$ at time $t$.

2. *Transform data so that higher values are associated with disease progression:* For each variable, we determine its 'direction.' If higher values of the variable are associated with greater disease severity then direction for the $v^{th}$ variable is given by $d_v = +1$, otherwise $d_v = -1$. We then define a new $N \times V \times M$ multi-dimensional array $Y$ such that $Y_{ivt} = d_v X_{ivt}$.

3. *Construct patient trajectory profiles:* For each patient $i$, we construct a $V$ x $M$ trajectory profile matrix, $T^i$. The matrix entries of $T^i$ are calculated as follows:
   - For non-binary variables:

     $$T^i_{vt} = 1 \quad if\ Y_{ivt} > \theta_v \qquad (1)$$
     $$= 0 \quad \text{otherwise}$$

     where $Y_{ivt}$ is defined as described above and $\theta_v$ is the threshold for the continuous variable $v$ that corresponds to the 50th percentile or median for $Y$ across the population in the baseline (t=0 for our data).For non-binary variables

   - For binary variables:
     *For gender:* $T^i_{vt}$ = 1 if the patient is male, $T^i_{vt}$ = 0 otherwise
     *For genetic risk loci:* $T^i_{vt}$= 1 if patient contains single nucleotide polymorphisms (SNP) $v$; $T^i_{vt}$ = 0 otherwise. Each SNP is treated as independent.

4. *Create a patient-patient network with connections based on trajectory similarity:* After having defined the trajectory profile matrix $T^i$ for each individual $i$, we create a patient-patient network $P$ of all patients in the training set with an adjacency matrix given by:

   $$P_{ij} = \sum_{v,t} (T^i_{vt} == T^j_{vt}). \qquad (2)$$

   In other words, $P_{ij}$ gives the number of matrix entries for which $T^i$ has the same value as $T^j$. This formulation implies that all variables are equally important. If we wanted to have different weights across variables and across the years, we can instead set $P_{ij} = \sum_{v,t} w_{vt}(T^i_{vt} == T^j_{vt})$ where $w_{vt}$ is the weight of variable $v$ at time $t$.

5. *Cluster the network to identify communities/subtypes:* We then perform Louvain community detection[18] to maximize the Newman-Girvan modularity function[19] on the uni-partite network defined by the weighted matrix $P$. This allows us to cluster trajectory profiles, and hence patients, into communities (subtypes).

6. *Construct aggregate profiles to characterize each community/subtype:* We average the trajectory profiles of all patients in each community $C^l$ to obtain the 'community/ subtype profile' $S^l$, a V x M matrix with elements defined by

   $$S^l_{vt} = \frac{\sum_{i \in C^l} T^i_{vt}}{N_l U_{v0}} \qquad (3)$$

   where $N_l$ is the total number of individuals in community $C^l$. $U_{v0}$ is a normalization constant that represents the average value for variable $v$ in the baseline:

$$U_{v0} = \frac{\sum_i T^i_{v0}}{N}$$, and again, 0 denotes the baseline year.

**Prediction Scheme for Test Patients**

From baseline data, we predict the community/subtype that an individual test patient (patient whose data was not used in identifying the PD subtypes) belongs to. We then check whether the test patient is still aligned with the same community/subtype after 4 years to demonstrate the utility of our baseline prediction.

To predict test patient *i*'s subtype from his/her baseline profile, we find the community (subtype) $C^l$ whose baseline community profile, with elements $S^l_{v0}$, has the smallest Euclidean distance from the patient's baseline profile. In other words, *l* is chosen to minimize the distance

$$d_0^{il} = \sqrt{\sum_v (T^i_{v0} - S^l_{v0})^2} \quad (4)$$

Does the patient's trajectory match the subtype's trajectory? We then investigate the quality of the subtype/community baseline prediction at a later time *t* by calculating the patient's subtype/community $C^{l'}$ such that *l'* is chosen to minimize the distance between the community profile and the patient's profile at time *t*:

$$d_t^{il'} = \sqrt{\sum_v (T^i_{vt} - S^{l'}_{vt})^2} \quad (5)$$

The prediction accuracy is then defined as the fraction of test patients for which the subtype identification (*l*) from the baseline matches the subtype identification (*l'*) at a later time *t.*

**TPC Algorithm for PD Subtype identification**

In this section, we present the disease subtypes (communities) identified by our method from the training patient data. Maximizing Newman-Girvan modularity on the patient-patient trajectory profile network gives us three distinct subtypes.

The darkness of the shade of grey of a continuous variable in a year denotes the fraction of the subtype population that has a value above the median of the total population baseline for that variable. The darkness of the shade of a grey for a binary variable is the fraction of the subtype population containing that variable (male in the case of the variable gender).

For some scales or tests, a higher score implies a healthier/less severely affected patient (such as the *Montreal Cognitive Assessment*), while for other scales, the opposite is true (higher score = greater severity). Therefore, in step one of our algorithm we normalized the data, so that for all variables except for the genetic and demographic variables, a higher score is associated with greater severity of that variable and a deeper shade of grey.

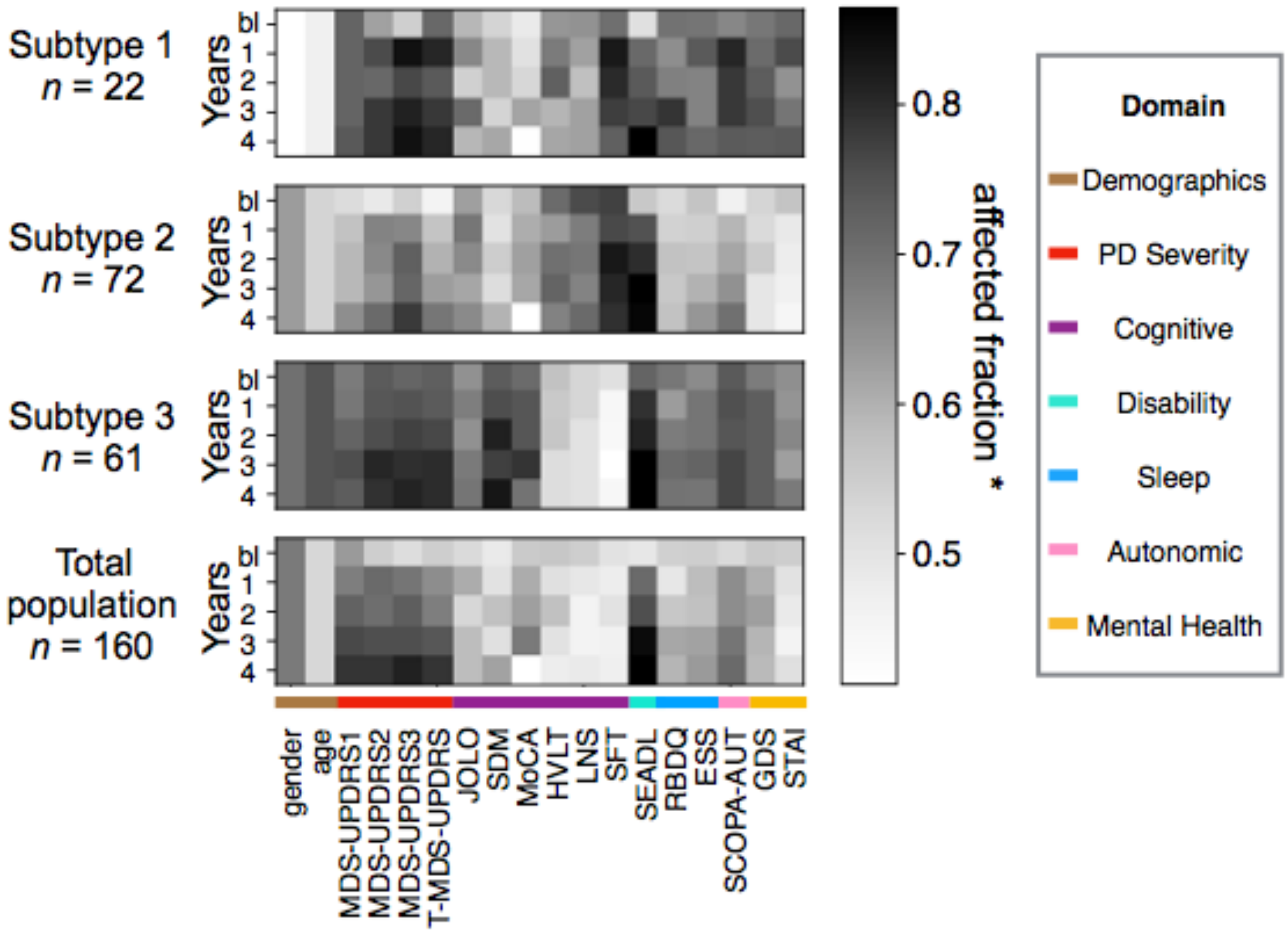

Figure 2: Subtypes/communities identified by our algorithm: The top three panels show three subtype/community profiles (average profile of all patients in the subtype). Subtypes identified by the algorithm containing fewer than 10 patients are not shown. The bottom panel shows the total population profile. The shade of grey indicates the affected fraction, i.e, fraction above baseline median in the direction of disease progression for the continuous variables, and fraction that is male for gender. *n* is the number of patients in the subtype. The variable names are listed below the panels (See Table 1 for description).

**Description of the Subtypes**

Subtype 1 is small (n=22), largely female and relatively young. Subtype 2 is the largest (n=72) with a milder disease course on measures of general PD severity, mental health, autonomic function and sleep, and a similar cognitive profile as Subtype 1. Subtype 3 shows a clearly different cognitive profile with better scores on some cognitive variables (SFT, HVLT, LNS), but worse scores on others (JOLO, SDM, MoCA). Subtype 3 is fairly large (n=61), largely male and older, with greater severity in variables describing general PD severity, sleep, autonomic function and mental health. Interestingly, the six cognitive variables split into two sub-categories (*Cog I:* JOLO, SDM, MoCA and *Cog II:* SFT, HVLT, LNS) especially in subtypes 2 and 3. The bottom panel in Figure 2 shows the profile of the total population. Since the threshold variable severity in an individual is set with respect to the median of the total population at baseline, the total population baseline profile for all variables has a value close to 0.5 (i.e., 50% of the total population at baseline has a value of 1 for any variable, and the other half has a value of 0). Fluctuations of the baseline total population value around 0.5 may occur if the precise value of a variable in the baseline year for multiple people coincides with the variable median of the entire population.

### Early Prediction of Patient Subtypes

In addition to identifying PD subtypes, our method predicts the individual patient subtype years in advance. In this section the test patient cohort (n=38) is used to assess the accuracy of early prediction of disease subtype. Data from these test patients is not used in the identification of the subtypes.

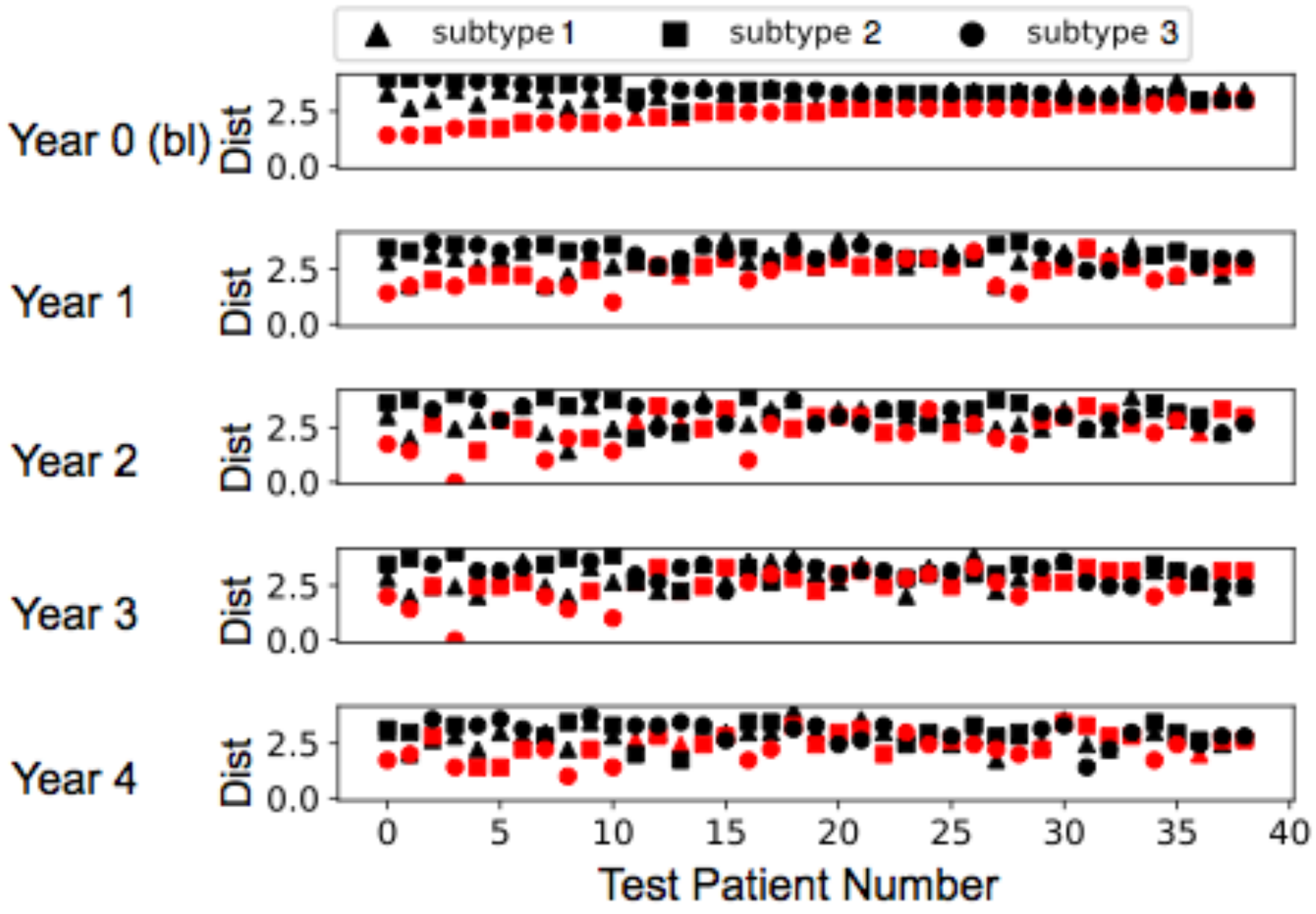

Figure 3: Prediction of test patients into the subtypes identified in Figure 2. The $i^{th}$ panel (row) shows the distance between the test patient $i^{th}$ year profile and the $i^{th}$ year subtype

profile (shape coded). The predicted subtype for each individual (subtype with minimum baseline-year distance) is colored red to allow for tracking across the years (panels). Prediction accuracy in year 4 is 74%. Data includes 38 test patients and 18 clinical variables across 5 time points: baseline (bl) or year 0 + years 1,2,3,4.

Figure 3 shows the prediction of future PD subtype based on baseline data for 38 test patients that run across the horizontal axis. The top panel shows the Euclidian distance between the baseline profile of a patient and the baseline profile of each subtype (subtypes are shape coded). The subtype with which the patient has minimum baseline distance is the 'predicted subtype', and is marked in red. Patients are organized from left to right in order of decreasing confidence, i.e., from minimum to maximum distance of the patients' baseline profile with the predicted subtype baseline profile. The remaining panels represent subsequent years, plotting the distance between the patient profile and subtype profiles in corresponding years. The red coding of predicted community makes it easy to track across the years. Finally, in year 4, we assess the accuracy of our predictions. If the subtype with minimum distance to the patient in year 4 is the same as the originally predicted subtype, then we consider our prediction to be successful for that patient. In other words, if in year 4 the red subtype for a patient is below the black subtypes then our prediction is successful. For the newly diagnosed PD patients in the PPMI dataset, our algorithm predicts the PD subtype after 4 years of disease progression with 74% accuracy.

**Incorporating Genetic Data into the TPC Algorithm**

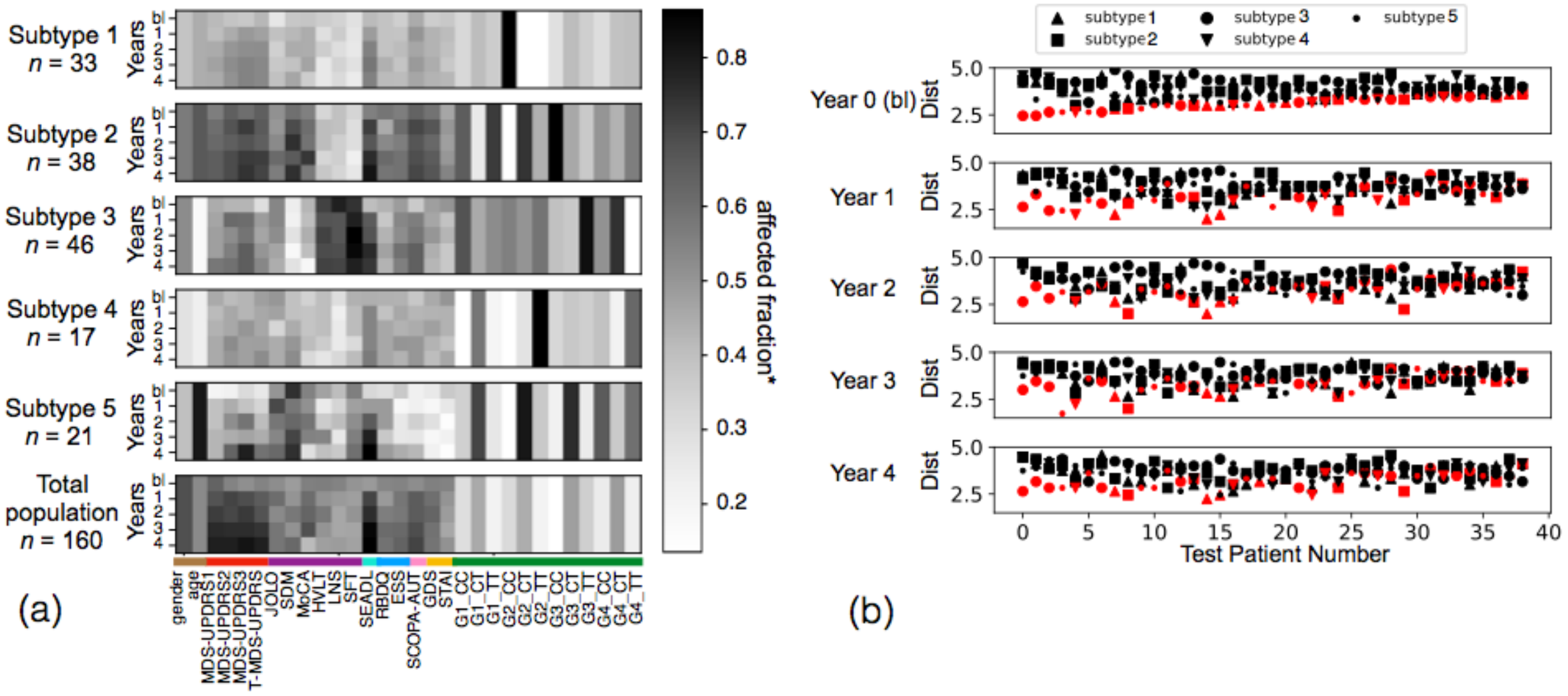

Figure 4: (a) The top five panels show three average community (subtype) profiles $C$ normalized by the total population baseline year (year 0), identified by our TPC algorithm. Subtypes identified by the algorithm containing less than 10 patients are not shown. The bottom panel shows the total population profile. The legend is a measure of the affected fraction, i.e, fraction above baseline median in the direction of disease progression for the continuous variables, and fraction that is male and fraction containing the genetic SNP for gender and genetic variables respectively. *n* is the number of patients in the community. (b) The $m^{th}$ panel shows the distance between the test patient $m^{th}$ year profile and the $m^{th}$ year profile of the subtypes (shape coded). The predicted subtype for each individual (subtype with minimum baseline distance) is colored red to allow for tracking across the years (panels). Prediction accuracy in year 4 is 64%. Data includes 38 test patients and 18 clinical variables across 5 time points: baseline (bl) or year 0 + years 1,2,3,4.

Genetic variants are increasingly recognized as important determinants of disease subtype and disease progression and prognosis. As an exploratory objective, we investigated the integration of genetic variants (single nucleotide polymorphisms, or SNPs) in previously identified PD risk loci into our TPC-based approach. Each patient has 2 copies for each piece of genetic information, and there are by definition 2 variants for each SNP. Hence, there are 3 possible combinations of the 2 variants for each of the genetic risk loci. PPMI contains information for 28 such SNPs for each patient. We chose 4 of those 28 loci to be included in our approach as a proof-of-principal. For one of those loci (rs356181/2), an association with PD motor subtype has recently been described,[17] making this an obvious choice for our study. In the absence of any established genotype-phenotype correlation for any of the other risk loci, we picked the additional 3 SNPs based on their high minor allele frequency, so that all 3 possible combinations of the 2 genetic variants were present in sufficient numbers in our study population of 198 subjects.

Figure 4 shows the five subtypes identified when genetic data is introduced. The plots Figure 4 (a,b) are organized in the same way as Figures 2 and 3 respectively. In Figure 4(a), the darkness of the shade of grey of a variable in a year denotes the fraction of the subtype that has a value above the baseline median of the total population of that variable. Subtypes 1 (top) and 4 have relatively mild disease profiles, with the CC allele of G2 being more frequent in subtype 1 (top), and the CT allele of G1, TT allele of G2, and TT allele of G4 being frequent in community 4. Subtypes 2 and 3 have alternate sets of cognitive variables dominating, a feature also observed in Figure 2. Subtypes 3 and 4 are the youngest, whereas 2 and 5 are the oldest. Subtype 2 has a larger population (38) and is the subtype with the maximum fraction of patients that have above average severity of the variables (darker grey shade). Subtype 5 has less psychiatric symptomatology but intermediate motor (MDS-UPDRS3) and cognitive dysfunction. Each of the subtypes has a distinct genetic profile. When genetic data is added to the analysis, prediction of patient subtype in the test group shows an accuracy of 64%. This is 10% less accuracy of prediction than obtained when using only the clinical data.

## Discussion

Multidimensional clinical datasets are valuable resources that are not used to their full potential due to the analytic challenges of diverse biomarkers and outcome variables. We describe development of a method to identify disease subtypes based on the pattern of progression of multidimensional clinical data including demographics, clinical variables, and genetics. We then validate our method by measuring the accuracy of subtype prediction in individual patients based on baseline clinical and genetic variables.

The disease subtypes are characterized by patterns of progression of the clinical variables. The concordance between our agnostic results with the domain-structure of the variables supports our approach. For instance, in the analysis on identifying PD subtypes, in spite of treating the variables independently, all the variables in the General PD severity domain show similar trends in each subtype. For example, subtypes 1 and 3 have high progression of all PD severity variables and subtype 2 has a low progression of all PD severity variables. Variables of other domains such as Sleep, Mental Health and Cognition show similar trends within a subtype (although Cognitive shows an additional layer of differentiation into Cog I and Cog II).

Our predictions of the future subtype of individual patients in the test sample based on their baseline data, shows good accuracy in predicting disease subtypes four years later (74% for clinical data and 64% for clinical+genetic data). The explanation for the reduction in predictive accuracy with addition of genetic data may be due to: 1) the inclusion of a very limited number of genetic risk loci, 2) that SNPs are more potent predictors of risk in PD than the clinical phenotype, 3) that genetic data isn't predictive of PD subtype within the 4-year time frame of our data or 4) that the genetic data has a large variance in the population, thus requiring a larger dataset for long-term prediction (the larger number of subtypes found by our method may indicate this). Nonetheless, this exploratory work successfully demonstrates the inclusion of genetic data in this approach. Other biomarkers (i.e. serologic and cerebrospinal fluid biomarkers) can also be easily integrated into our analysis. Our algorithm is likely to benefit from more extensive datasets with larger populations.

A number of studies have identified PD subtypes based on baseline characteristics[20-22]. In this work, we used the longitudinal data to identify disease subtypes, and defined the baseline characteristics of the subtypes retrospectively. The baseline features of individual patients were then used in a test cohort to predict the future disease trajectory (prognosis). To our knowledge, this is a novel approach[21]. Our study represents an innovative approach, that has advantages over previous methods by taking full advantage of large heterogenous, longitudinal datasets.

Our trajectory clustering method works with various types of data including clinician- and patient-reported outcome measures, genetic alleles, physical performance measures, as well as diverse results from diagnostic investigations. This first approach uses demographics, clinician- and patient-reported data, and genetic data. In this analysis, each genetic SNP and clinical variable is treated independently and allotted the same weight. Our algorithm allows for variable weightings, where each domain and SNP is assigned a chosen weight. However, this raises the question of how the weighting would be decided. For example, if we had allotted equal weights to *one hundred SNPs* in our analysis in addition to the 18 clinical variables, the genetic information would dominate the algorithm, and affect the resulting communities. On the other hand, different weighting strategies may be preferable based on the study aims. For example, if the main objective is to identify disease subtypes based on motor vs. cognitive function, one could allot equal cumulative weight to the motor and cognitive domains.

A strength of our algorithm, which is also a caveat, is that it is entirely data-driven. The level of severity of each variable relative to the baseline median is used to normalize all variables, as opposed to the absolute value of the variable. This is done so as to readily compare changes in different variables. A notable example is the clinical variable, SEADL (a disability scale). SEADL is a relatively insensitive scale in the early years of PD since there is little functional disability in the years following diagnosis. Yet, in our analysis SEADL shows high progression (darker shade in later years) in Figures 2 and 4(a). It is important to note that this dark shade isn't indicative of the absolute severity. It only tells us that a larger fraction of the population in the later years has SEADL values above the baseline mean (which may be low to begin with).

Our approach is innovative, adaptable, and clinically relevant. PD subtyping[24] is an area of active research but there are currently no clinically prognostic analyses in use in the management of PD. Reliable prognostic analyses will change clinical management by informing earlier, more aggressive management and improving prognostic counseling. A natural extension of this work involves implementing the method for datasets in other chronic medical conditions. Another interesting direction involves extending the TPC algorithm to incorporate and compare other network clustering approaches, such as multi-layer network clustering[25]. Other future directions include studies of the effect of treatment on progression of disease variables, and predicting modifications of algorithm-identified subtypes as a consequence of different treatments.

**Acknowledgement**

This work was supported in part by the UMB-UMD Research and Innovation Seed Grant Program. S.K. and M.G. also received support for their contributions through NSF award DGE-1632976.